\documentclass[aip,jcp,numerical,preprint]{revtex4-2}

\usepackage{amsmath}
\usepackage{epstopdf}
\usepackage{graphicx}
\usepackage{dcolumn}
\usepackage{bm}
\usepackage{braket}
\usepackage{gensymb}
\usepackage{array}
\newcolumntype{P}[1]{>{\centering\arraybackslash}p{#1}}
\newcolumntype{M}[1]{>{\centering\arraybackslash}m{#1}}

\newcommand{\bs}{\boldsymbol}

\newcommand{\ig}{\text{i}}

\newcommand{\ex}{\text{e}}

\begin{document}

\title{Measuring Time-Dependent Induced Quantum Coherences  via Two-Dimensional  Coherence Spectroscopy}

\author{William Barford}
\email{william.barford@chem.ox.ac.uk}
\affiliation{Department of Chemistry, Physical and Theoretical Chemistry Laboratory, University of Oxford, Oxford, OX1 3QZ, United Kingdom}

\author{Allison Nicole Arber}
\affiliation{Department of Chemistry, Physical and Theoretical Chemistry Laboratory, University of Oxford, Oxford, OX1 3QZ, United Kingdom}
\affiliation{Balliol College, University of Oxford, Oxford, OX1 3BJ, United Kingdom}

\author{Fynn McLennan}
\affiliation{Department of Chemistry, Physical and Theoretical Chemistry Laboratory, University of Oxford, Oxford, OX1 3QZ, United Kingdom}
\affiliation{Magdalen College, University of Oxford, Oxford, OX1 4AU, United Kingdom}

\author{Max Marcus}
\affiliation{Department of Chemistry, Physical and Theoretical Chemistry Laboratory, University of Oxford, Oxford, OX1 3QZ, United Kingdom}

\begin{abstract}
We propose a two-dimensional spectroscopic  protocol for measuring the time-dependent coherences between the stationary states of a system induced by a time-dependent system-bath interaction. We also investigate the role of temporally-correlated noise on coherence dephasing. This protocol enables  dynamical information about the system and its coupling to the environment to be determined.
Our results are based on the quantum-trajectory method, and are obtained from both  approximate, analytical and exact, numerical solutions of the time-dependent Schr\"odinger equation.
As an example, we show how this protocol can be used to investigate exciton dynamics in conjugated polymers induced by the coupling of their torsional modes with the environment.
\end{abstract}

\maketitle


\section{Introduction}\label{Se:1}

The question of whether electronic and vibrational  coherences exist in
macromolecular systems, e.g.,\ $\pi$-conjugated polymers and light harvesting complexes, has been an outstanding one for a number of years\cite{Engel07,Fleming09,Cheng09,Scholes11a,Scholes11b,Lewis12,Scholes15,Scholes17}. This question is  motivated  by the attempt to understand how coherences --- if they exist --- can survive under ambient conditions. It is also motivated by the expectation that coherences might enhance the efficiency of energy and charge transport (but see Ref.\ \cite{Kassal13} for a counter argument), thus improving the efficiency and providing design principles for synthetic photovoltaic devices.

Recently, borrowing concepts from quantum information theory, quantum process tomography has  been proposed as a method to `witness'
coherences\cite{Yuen-Zhou11,MMarcus18,MMarcus20}.
In this paper we propose a conceptually  straightforward protocol based on the well-established technique of two-dimensional
coherence spectroscopy  to determine dynamical coherences. For reviews of two-dimensional coherence spectroscopy, see Refs\cite{Jonas03,Cho08,Cheng09,Hamm11,Lewis12}.

Before discussing how coherences are established and measured, let us first define what we mean by them. In this paper we define
coherences  as the off-diagonal matrix elements of the  system's density matrix when expressed in the energy eigenbasis of the system. Such coherences in a system can be induced via coupling to the environment in a variety of ways. For example, a coherent light source might excite a number of energy eigenstates, thus creating a non-stationary state. In this case, assuming no dephasing or dissipation, the populations are constant, while  coherences have constant magnitudes but oscillate with the transition angular frequencies. A more interesting scenario is when the excited system interacts with an environment that couples these eigenstates and hence causes interstate transitions. This will cause population transfers and the magnitude of the coherences will change in time. Moreover, if the energy eigenstates are spatially separated, it will cause energy transfer. It is this latter scenario that is the principal investigation of this paper.

Here, we set up a simple two-level system subject to a time-dependent periodic interaction. We show how the induced coherences can be observed by using 2D-spectroscopy with particular choices of the electric field polarization. This results in a characteristic fingerprint of dynamical coherences, from which  dynamical information about the system and environment can be determined. We then demonstrate how a noisy (dephasing) environment destroys the coherences. Our results are obtained from both  approximate, analytical and exact, numerical solutions of the time-dependent Schr\"odinger equation.

Having established this model system, we next discuss how it might be realized in practice. In particular, we envisage a conjugated polymer subject to torsional fluctuations via the Brownian impulses of its environment. We derive realistic parameters for both intra and inter chromophore energy transfer, and investigate whether the coherence signals described in Section \ref{Sec:2.4}  might be observed.

\section{Two-Level System}\label{Se:2}

\subsection{Model}\label{Se:2.1}

To introduce the key ideas behind the spectroscopic technique proposed here, we  begin by considering a closed two-level system described by the Hamiltonian,
\begin{equation}
\hat{H}_{\text{S}} = \hbar\omega_a \ket{a}\bra{a} + \hbar\omega_b\ket{b}\bra{b}.
\end{equation}
Note that the true ground state, $\ket{\text{GS}}$, is not part of this model, as the kets $\ket{a}$ and $\ket{b}$ span the excited state Hilbert space. Consequently, the energies $\hbar\omega_j$ are excitation energies, rather than absolute energies. Defining $\hat{a}_j^{\dagger}$ as the creation operator for ket $\ket{j}$, then $\ket{j}=\hat{a}_j^{\dagger}\ket{\text{GS}}$.
For convenience we also define the Bohr angular frequency $\omega_0 = (\omega_b - \omega_a) > 0$.

The system is coupled to a bath via the interaction
\begin{equation}\label{Eq:2}
  \hat{H}_{SB} = V(t)\left(\ket{a}\bra{b}+ \ket{b}\bra{a}\right),
\end{equation}
where we take the system-bath interaction, $V(t)$, to be of the general form
\begin{equation}\label{Eq:3}
 V(t) = 2\hbar \varpi \cos( \omega t + \phi(t)+ \chi).
\end{equation}
Here, $2\hbar\varpi$ is the interaction strength, $\omega$ is the driving frequency and $\phi(t)$ acts as a temporally-correlated phase that causes dephasing of the coherences. $\chi$ is a constant phase whose origin will be explained later.

An electric field pulse acting on the ground state will excite the system into these two states. We denote the electric field as
\begin{equation}
{\bf E}(t) = E(t){\bf e},
\end{equation}
where $E(t)$ is the temporal envelope of the pulse and ${\bf e}$ is its polarisation. If the pulse (centered at time $t=0$) is sufficiently narrow in time  it will  act instantaneously  on the system and thus the excited  state is given as,\cite{Tannor07, Yuen-Zhou14}
\begin{equation}
\ket{\Psi(t=0)} = \frac{\ig}{\hbar}\sum_{j=a,b} ({\bm \mu}_j \cdot{\bf e}) \tilde{E}(\omega_j)\ket{j} = \sum_{j=a,b} \psi_j(0)\ket{j},
\end{equation}
where $\bm{\mu}_{j} = \bra{j} \bm{\mu} \ket{\text{GS}}$ is the transition dipole moment and $\tilde{E}(\omega_j)$ is the Fourier transform of $E(t)$ at frequency $\omega_j$. At a subsequent time the system is subject to evolution as determined by the total Hamiltonian, $\hat{H}_{S}+\hat{H}_{SB}$. In general,
\begin{equation}\label{Eq:5}
\ket{\Psi(t)} = \sum_{j=a,b} \psi_j(t)\ex^{-\ig\omega_j t}\ket{j},
\end{equation}
where the amplitudes $\psi_j$ are time-dependent as the time-dependent interaction on the system  induces transitions between the eigenstates.


\subsection{Response Functions}

In this work we determine the third-order non-linear response functions using the method of quantum trajectories, as described  by Marcus \emph{et al.}\cite{AHMarcus07}. This wavefunction method is particularly convenient for determining numerical solutions in Hilbert space, as it avoids dealing with the much larger Liouville space that is necessary in a denisty matrix formalism.

The first order trajectory, $\ket{\Psi_{\beta}(t_{\delta})}$, is defined as an excitation from the ground state by a single pulse at $t = t_{\beta}$  and subsequent evolution to a time $t_{\delta} \geq t_{\beta}$. Thus,
\begin{equation}\label{Eq:70}
\ket{\Psi_{\beta}(t_{\delta})} = \hat{U}(t_{\delta},t_{\beta})\hat{O}_{\beta}\ket{\text{GS}},
\end{equation}
where $\hat{O}_{\beta}$ is an excitation operator defined for a given spectroscopic protocol and $\hat{U}$ is the evolution operator.
Similarly, the third order trajectory, $\ket{\Psi_{\alpha\gamma\delta}(t_{\delta})}$, is created by  three pulses at times $t_{\alpha} \leq t_{\gamma} \leq t_{\delta}$. Thus,
\begin{equation}\label{Eq:71}
\ket{\Psi_{\alpha\gamma\delta}(t_{\delta})} = \hat{O}_{\delta}\hat{U}(t_{\delta},t_{\gamma})\hat{O}_{\gamma}\hat{U}(t_{\gamma},t_{\alpha})\hat{O}_{\alpha}\ket{\text{GS}},
\end{equation}
again where the (de)excitation operators will be explained later.

\begin{figure}
\includegraphics[width=0.75\linewidth]{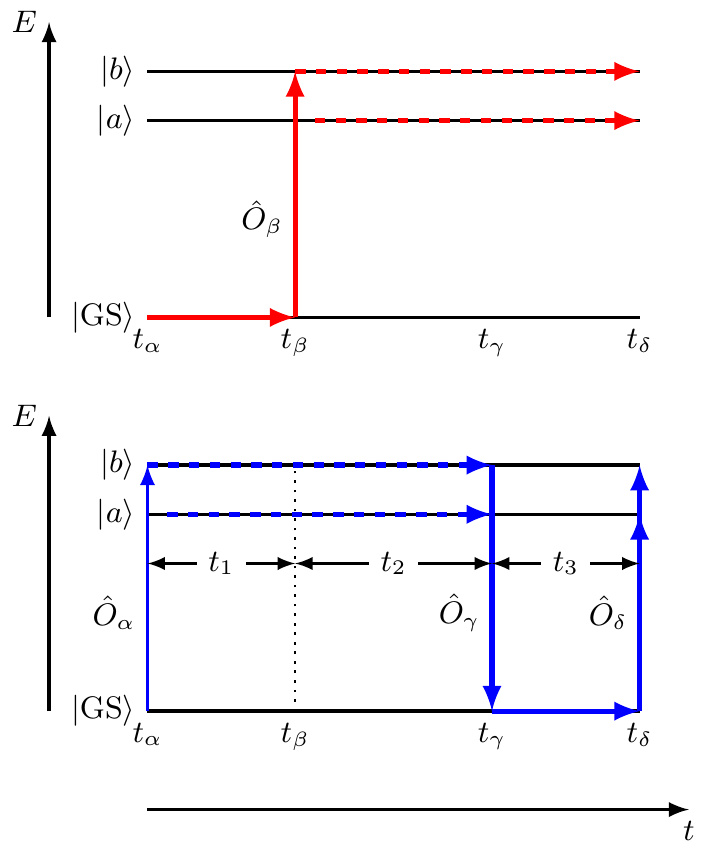}
\caption{Our proposed spectroscopic protocol for measuring coherences induced by an external interaction, as described in Section \ref{Sec:2.4}. The upper panel shows the evolution of the first-order trajectory, $\ket{\Psi_{\beta}}$, after an excitation into $\ket{b}$ and subsequent oscillation between $\ket{b}$ and $\ket{a}$. The lower panel shows the third-order trajectory, $\ket{\Psi_{\alpha\gamma\delta}}$, after an excitation into $\ket{b}$, subsequent  oscillation between $\ket{b}$ and $\ket{a}$, then de-excitation and re-excitation of the amplitudes of both  $\ket{a}$ and $\ket{b}$. The dashed horizontal lines indicate oscillations between $\ket{b}$ and $\ket{a}$.}
\label{Fig:1}
\end{figure}

The ordering of $t_{\beta}$ relative to $t_{\alpha}$, $t_{\gamma}$ and $t_{\delta}$ determines the type of diagram in 2D-spectroscopy, i.e., rephasing or non-rephasing\cite{Hamm11,AHMarcus07}. To measure dynamically-induced coherences a rephasing (or photon echo) diagram is required. We choose the `stimulated emission' diagram, illustrated in Fig.\ \ref{Fig:1}, where $t_{\alpha} \leq t_{\beta} \leq t_{\gamma}$.
\footnote{This diagram is labeled `B' in Ref.\cite{AHMarcus07} and $R_1$ in Ref.\cite{Hamm11}.}

The third-order response is then given by the overlap of these two trajectories, namely,
\footnote{Setting $\hat{O}_{\delta} \equiv \hat{\mu}$ gives $\left\langle\Psi_{\alpha\gamma\delta}|\Psi_{\beta}\right\rangle \equiv \textrm{Tr} \left\{\hat{\mu} \hat{\rho}^{(3)}\right\}$, where $\hat{\rho}^{(3)}$ is the third-order density operator\cite{Hamm11} for a pure state. Similarly, an ensemble average over $S(t_1,t_2,t_3)$ implies  that $\hat{\rho}^{(3)}$ is the third-order density operator for a mixed state. Indeed, Eq.\ (\ref{Eq:9a}) is the equivalent in third-order non-linear spectroscopy to the Loschmidt echo, $S(t) = \left\langle\Psi_{\alpha}|\Psi_{\beta}\right\rangle$, in linear spectroscopy\cite{Tannor07}.}
\begin{equation}\label{Eq:9a}
S(t_1,t_2,t_3) = \left\langle\Psi_{\alpha\gamma\delta}|\Psi_{\beta}\right\rangle.
\end{equation}
Defining the time variables as $t_1 = t_{\beta} - t_{\alpha}$ (coherence time), $t_2 = t_{\gamma} - t_{\beta}$ (waiting time) and $t_3 = t_{\delta} - t_{\gamma}$ (echo time), and Fourier transforming with respect to $t_1$ and $t_3$, we obtain the two-dimensional in frequency-space and one-dimensional in time-space third-order response,
\begin{equation}\label{Eq:15}
  \tilde{S}(\omega_1, t_2, \omega_3) = \int S(t_1,t_2,t_3)  \exp(\textrm{i}(\omega_1 t_1 + \omega_3 t_3))  \textrm{d}t_1 \textrm{d}t_3.
\end{equation}

Details of how the third-order signal is measured may be found in text books; see Hamm and Zani\cite{Hamm11}, for example.

\subsection{Closed System, $\hat{H}_{SB}= 0$}

To motivate our experimental protocol for measuring dynamically-induced coherences, as described in Section \ref{Sec:2.4}, we first consider two simple examples of determining a 2D-spectrum in a closed system, i.e., $\hat{H}= \hat{H}_S$.

\subsubsection{Stationary State}\label{Sec:2.1}

Let us first consider the case of a targeted excitation within the two-level system. We assume that the pulse at $t_{\beta}$ only  excites the system into the $\ket{b}$ state. In this case $\hat{O}_{\beta} = \hat{a}_b^{\dagger}$ and the first-order trajectory is,
\begin{equation}
\ket{\Psi_{\beta}} = \hat{U}(t_{\delta},t_{\beta})\hat{a}_b^{\dagger}\ket{\text{GS}} = \ex^{-\ig\omega_b (t_{\delta} - t_{\beta})}\ket{b}.
\end{equation}
Similarly, for the third-order trajectory with $\hat{O}_{\alpha} = \hat{O}_{\gamma}^{\dagger} = \hat{O}_{\delta} = \hat{a}_b^{\dagger}$, we obtain,
\begin{equation}
\ket{\Psi_{\alpha\gamma\delta}} = \ex^{-\ig\omega_b(t_{\gamma}-t_{\alpha})}\ket{b},
\end{equation}
as in the ground state the evolution operator is just the identity.

Then,
\begin{equation}
S(t_1,t_2,t_3)  = \ex^{\ig\omega_b(t_{\beta}-t_{\alpha}+t_{\gamma} - t_{\delta})} = \ex^{\ig\omega_b(t_1 - t_3)}
\end{equation}
and
\begin{equation}
\begin{split}
\tilde{S}(\omega_1,t_2,\omega_3) = \delta(\omega_3-\omega_b)\delta(\omega_1 + \omega_b).
\end{split}
\end{equation}
This shows that the resulting 2D spectrum consists of a single peak at $(-\omega_b,\omega_b)$ in the $(\omega_1,\omega_3)$ plane, which is independent of $t_2$. As, once excited the population of $\ket{b}$ does not change, this peak is a direct signal for the population of the investigated state. A schematic spectrum is shown in Fig.\ 2(a).

\begin{figure*}\label{Fig:2}
\centering
\includegraphics[width=0.95\linewidth]{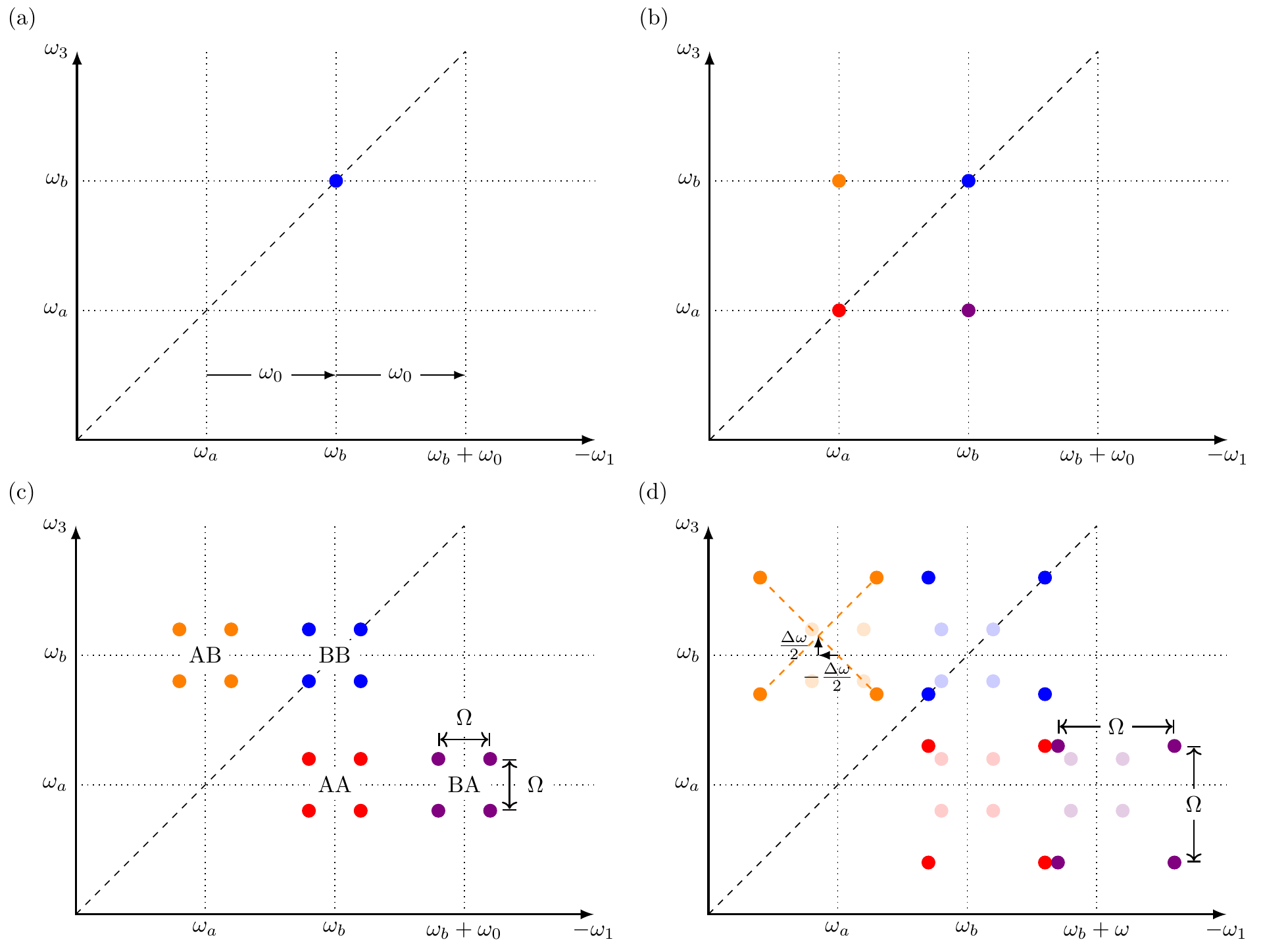}
\caption{Schematics of the analytical spectra of a two-level system. (a) shows a single excitation of a closed system (i.e., a stationary state) and  (b) a collective excitation into both states (i.e., a non-stationary state).
(c)  and (d) shows the analytical spectra, determined using the RWA (and given in Appendix A), of the system under the influence of a time-dependent interaction with the excitation sequence shown in Fig.\ 1. (c)  on-resonant interaction
while (d) is off-resonant. The red and blue signals correspond to the population signals of states $\ket{a}$ and $\ket{b}$, while the purple and orange signals are signatures of coherence between the states. The interaction shifts the signals at $\omega_3 = \omega_a$ to the right by  the driving frequency, $\omega$, and splits each signal into four by the Rabi frequency, $\Omega$. At off-resonance the detuning leads to a shift of the four groups of four with respect to the positions of the on-resonance peaks (shown faintly in (d)). The waiting time $t_2  = 0$.}
\end{figure*}

\subsubsection{Non-stationary State}

Let us now consider the 2D-spectrum obtained via the coherent simultaneous excitation of $\ket{a}$ and $\ket{b}$ and the non-stationary state's subsequent evolution under the action of $\hat{H} _S$.
In this case,
\begin{equation}
\hat{O}_{\alpha} = \hat{O}_{\beta} = \hat{O}_{\gamma}^{\dagger} = \hat{O}_{\delta} = \frac{1}{\sqrt{2}}\left(\hat{a}_a^{\dagger}+\hat{a}_b^{\dagger}\right).
\end{equation}
The first-order trajectory  is then
\begin{equation}
\ket{\Psi_{\beta}} = \frac{1}{\sqrt{2}}\left(\ex^{-\ig\omega_a(t_{\delta}-t_{\beta})}\ket{a}+\ex^{-\ig\omega_b(t_{\delta}-t_{\beta})}\ket{b}\right),
\end{equation}
while the third-order trajectory  is
\begin{equation}
\ket{\Psi_{\alpha\gamma\delta}} = \frac{1}{2\sqrt{2}}\ex^{-\ig\omega_a(t_{\gamma}-t_{\alpha})}\left(1+ \ex^{-\ig\omega_0(t_{\gamma}-t_{\alpha})}\right)\left(\ket{a}+\ket{b}\right).
\end{equation}
The  time-resolved overlap is
\begin{widetext}
\begin{equation}
S(t_1,t_2,t_3) = \frac{1}{4}\left[ \ex^{-\ig\omega_a(t_3-t_1)} + \ex^{-\ig\omega_b(t_3-t_1)}+ \ex^{-\ig\omega_a(t_3-t_1)} \left(\ex^{-\ig\omega_0(t_3+t_2)} + \ex^{\ig\omega_0(t_2+t_1)}\right)\right],
\end{equation}
\end{widetext}
while the frequency-resolved spectrum is,
\begin{widetext}
\begin{equation}
\begin{split}
\tilde{S}(\omega_1,t_2,\omega_3) &= \frac{1}{4}\left[\delta(\omega_1+\omega_a)\delta(\omega_3-\omega_a) + \delta(\omega_1+\omega_b)\delta(\omega_3-\omega_b)\right.\\& +\left. \left(\delta(\omega_1+\omega_a)\delta(\omega_3-\omega_b)\ex^{-\ig\omega_0t_2}+\delta(\omega_1+\omega_b)\delta(\omega_3-\omega_a)\ex^{\ig\omega_0t_2}\right)\right].
\end{split}
\end{equation}
\end{widetext}

Here we see that the resulting spectrum has four peaks: two are on the diagonal of the $(-\omega_1,\omega_3)$ axes, representing the populations of the two eigenstates, while the other two are  off-diagonal and oscillate with a frequency $\omega_0$ in  $t_2$.  These off-diagonal peaks are the coherences, $\psi_a^*\psi_b$, between the two eigenstates.  A schematic spectrum is shown in Fig.\ 2(b).

We see that if both states are populated the spectrum becomes richer. As the system here is closed, the states have to be populated directly by the electric field pulses, as otherwise no coherences occur. However, if the system is subject to an interaction and population transfer can happen during its evolution, then the initial excitation can be into a single state, thereby probing the possibility of population (and hence energy) transfer. We  discuss this situation in the next section.

\subsection{Open System with a Time-Dependent Interaction,  $\hat{H}_{SB}$}\label{Sec:2.4}

We now turn to the key result of this work: a protocol for measuring the coherences of a system induced by a time-dependent interaction. In this case we assume that both the first and third order trajectories are initially prepared in one of the eigenstates of $\hat{H}_{S}$. We choose this to be the upper state,  $\ket{b}$. Then, under the action of $\hat{H}_{SB}$ the system oscillates coherently between $\ket{b}$ and $\ket{a}$. Specifically, the amplitudes in Eq.\ (\ref{Eq:5}) are given by the time-dependent Schr\"odinger equations,
\begin{equation}
\frac{\textrm{d}\psi_a(t)}{\textrm{d}t} = -\ig2\varpi\cos(\omega t+ \phi(t)+\chi)\ex^{-\ig\omega_0 t}\psi_b(t)
\end{equation}
and
\begin{equation}
\frac{\textrm{d}\psi_b(t)}{\textrm{d}t} = -\ig2\varpi\cos(\omega t+\phi(t)+ \chi)\ex^{\ig\omega_0 t}\psi_a(t).
\end{equation}

These equations can be solved within the rotating-wave or secular approximation (RWA), which  assumes that $(\omega + \omega_0) \gg |\omega - \omega_0|$.
Using the method of Laplace transforms, with the initial conditions of $\psi_a(0) = 0$ and $\psi_b(0) = 1$, and setting the temporal phase $\phi(t)=0$, we obtain
\begin{widetext}
\begin{equation}\label{Eq:8}
  \psi_a(t) = \left( \frac{\varpi}{\Omega} \right) \exp(\textrm{i}  \chi)\exp(\textrm{i} \Delta \omega t/2 ) \left(\exp(\textrm{i} \Omega t/2)-\exp(-\textrm{i} \Omega t/2)\right)
\end{equation}
\end{widetext}
and
\begin{widetext}
\begin{equation}\label{Eq:9}
  \psi_b(t) = \frac{1}{2} \exp(-\textrm{i} \Delta \omega t/2) \left(\left(1+\left(\frac{\Delta \omega}{\Omega}\right)\right)
  \exp(\textrm{i} \Omega t/2)+\left(1-\left(\frac{\Delta \omega}{\Omega}\right)\right)\exp(-\textrm{i} \Omega t/2)\right).
\end{equation}
\end{widetext}
The detuning parameter is
\begin{equation}\label{Eq:10}
 \Delta \omega = (\omega - \omega_0)
\end{equation}
and we define the Rabi angular frequency as
\begin{equation}\label{Eq:11}
  \Omega = (4\varpi^2 +(\Delta \omega)^2)^{1/2}.
\end{equation}

The time-dependent system-bath interaction causes population to be exchanged between $\ket{b}$ and $\ket{a}$ with a Rabi time period $T_{\Omega} = 2 \pi/\Omega$.
For the pure quantum state defined by Eq.\ (\ref{Eq:5}) the populations of the eigenkets are $|\psi_a(t)|^2$ and $|\psi_b(t)|^2$, while their coherences are $\psi_a^*(t)\psi_b(t)$.
Off-resonance (i.e., $\Delta \omega \neq 0$) there is an incomplete transfer of population with the maximum population of  $\ket{a}$ being $4({\varpi}/{\Omega})^2$.

We notice the phase factor of $\exp(\textrm{i}  \chi)$ in the expression for $\psi_a(t)$, which arises from the $t=0$ phase in $V(t)$. As explained shortly, this phase factor has important implications for the definition of $\ket{\Psi_{\beta}}$, which is created at a time $t_1 = (t_{\beta} - t_{\alpha})$ after $\ket{\Psi_{\alpha}}$ at which point $\chi = \omega t_1$.

We now propose a 2D-spectroscopic protocol for observing these induced coherences.
The first-order  trajectory $|\Psi_{\beta}\rangle$ is given by Eq.\ (\ref{Eq:70}),
where $\hat{O}_{\beta} = \hat{a}^{\dagger}_b$ and $\hat{U}(t_{\delta},t_{\beta})$ is  now determined by $\hat{H}_S$ and $\hat{H}_{SB}$.
Thus, $|\Psi_{\beta}\rangle$ is an excitation from the groundstate to $\ket{b}$ at time $t_{\beta}$, followed by evolution in the excited state manifold (ESM) under the action of $(\hat{H}_S+\hat{H}_{SB})$ to time $t_{\delta}$, by which time it has acquired amplitude in both  $\ket{a}$ and  $\ket{b}$. This trajectory is illustrated schematically in red in Fig.\ \ref{Fig:1}.

Similarly, the third-order trajectory $\ket{\Psi_{\alpha\gamma\delta}}$ is given by Eq.\ (\ref{Eq:71}),
where $\hat{O}_{\alpha} = \hat{a}^{\dagger}_b$, but now $\hat{O}_{\gamma} = (\hat{a}_a+\hat{a}_b)/\sqrt{2}$ and $\hat{O}_{\delta} = (\hat{a}^{\dagger}_a+\hat{a}^{\dagger}_b)/\sqrt{2}$.
Thus, $\ket{\Psi_{\alpha\gamma\delta}}$ is (i) an excitation from the ground state to $\ket{b}$ at time $t_{\alpha}$, (ii) evolution in the ESM under the action of $(\hat{H}_S+\hat{H}_{SB})$ to time $t_{\gamma}$, (iii) de-excitation via a linear combination of $\ket{a}$ and $\ket{b}$ to the ground state  at time $t_{\gamma}$,  (iv) evolution in the ground state to time $t_{\delta}$, and finally (v) re-excitation via a linear combination of $\ket{a}$ and $\ket{b}$ at  time $t_{\delta}$. This trajectory is illustrated schematically in blue in Fig.\ \ref{Fig:1}. The selective (de)excitation of $\ket{a}$ and $\ket{b}$ is achieved by different polarizations of the laser light, as illustrated in Section III.

By a  time $t_{\gamma}$ the state $\ket{\Psi_{\alpha}} = \hat{U}(t_{\gamma},t_{\alpha})\hat{O}_{\alpha}|\textrm{GS}\rangle$ has acquired amplitude in both  $\ket{a}$ and  $\ket{b}$. These amplitudes are transferred to the ground state via $\hat{O}_{\gamma}$ and re-excited at $t_{\delta}$ via $\hat{O}_{\delta}$. The interference of these amplitudes with those of $|\Psi_{\beta}\rangle$, via the overlap $\langle\Psi_{\alpha\gamma\delta} |\Psi_{\beta}\rangle$, allow the dynamics --- and in particular, the coherences --- of the system to be determined.

Since the wavefunction overlap, $\langle\Psi_{\alpha\gamma\delta} |\Psi_{\beta}\rangle$, obtained via the RWA (i.e., Eq.\ (\ref{Eq:8}) and (\ref{Eq:9})) is a sum of a product of complex exponentials, its Fourier transform with respect to $t_1$ and $t_3$ results in a sum of a product of delta-functions. In particular, $\tilde{S}(\omega_1, t_2, \omega_3)$ corresponds to four groups of four-peaks whose analytical expressions are given in Appendix A. We now describe the key features of the resulting 2D-spectrum for cases of  on-resonance and off-resonance interactions.

\subsubsection{On-Resonance, $\omega = \omega_0$}


At resonance there is complete population transfer between $\ket{b}$ and $\ket{a}$. The resulting spectrum is given in Appendix A and  illustrated by Fig.\ 2(c). From this we observe:
\begin{enumerate}
\item{As explained in Section \ref{Sec:2.1}, in the absence of induced coherences, i.e., when $V(t) = 0$, the system would remain in the stationary state $\ket{b}$ and there would be a single population peak at $(-\omega_b,\omega_b)$ (as in Fig.\ 2(a)).}
\item{With induced coherences, however, there are four groups of four peaks. These groups of peaks correspond to the populations of $\ket{b}$ (i.e., at BB) and  $\ket{a}$ (i.e., at AA), and their coherences (i.e., AB and BA).}
\item{Notice, however, that the two groups of peaks at $\omega_3 = \omega_a-\Delta \omega/2$ (i.e., AA and BA) are displaced along the $-\omega_1$ axis by $\omega$.
This `boost' along $\omega_1$ at $\omega_3 = \omega_a-\Delta \omega/2$ is a consequence of the phase factor $\chi = \omega t_1$ that the amplitude of $\ket{a}$ acquires when $\ket{\Psi_{\beta}}$ is created, as shown by Eq.\ (\ref{Eq:8}).}
\item{Each group of peaks is split into four peaks whose splitting is determined by $\Omega=2\pi/T_{\Omega}$, where $T_{\Omega}$ is the (Rabi) population transfer period.}
\item{Items (3) and (4)\ illustrate the characteristic 2D-spectrum obtained via this 2D-protocol that provides a characteristic fingerprint of dynamically-induced  coherences.}
\item{The peaks exhibit complex dynamics as a function of $t_2$. For example, the off-diagonal components of the population group of peaks oscillate with a period of $T_{\Omega}$, while the diagonal components of the coherence group of peaks oscillate with a period of $T = 2 \pi/\omega$.}
\end{enumerate}

\subsubsection{Off-Resonance, $\omega \neq \omega_0$}

\begin{figure}
\includegraphics[width=0.8\linewidth]{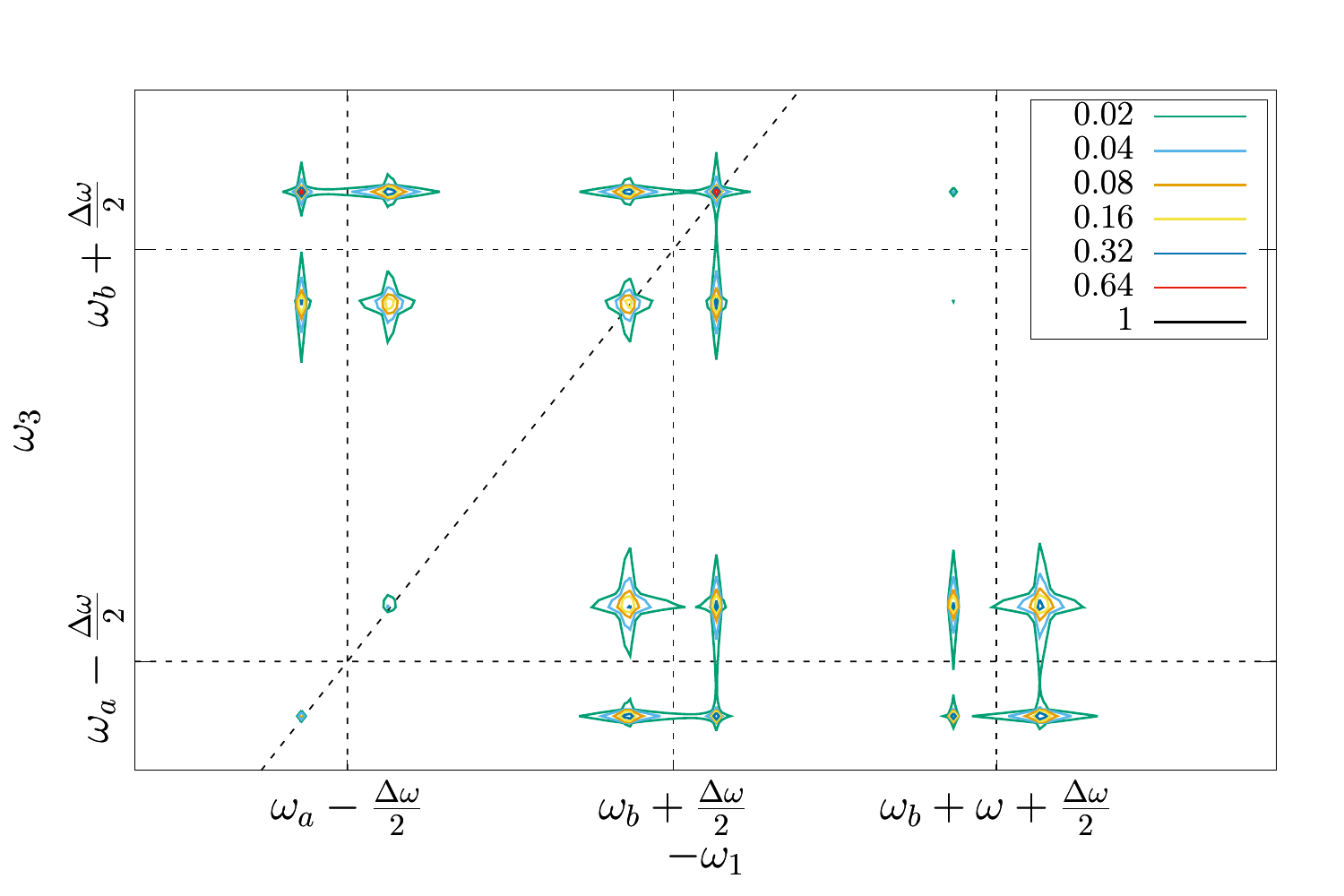}
\caption{The 2D-spectrum for $\Delta\omega/\omega_0 = 0.9$, $\varpi/\omega_0 = 0.25$ and $t_2=0$. These results are obtained by computing the overlap $\langle\Psi_{\alpha\gamma\delta} |\Psi_{\beta}\rangle$ numerically using the TNT Library\cite{Al-assam17} and evaluating Eq.\ (\ref{Eq:15}) using a discrete Fourier transform (DFT).
This should be compared to the analytical solution obtained via the RWA, shown in Fig.\ 2(d).
Note, however, that additional  low-intensity  peaks  appear here because this an exact numerical calculation of the wavefunction overlap.
The peak features (i.e., elongations along the axes) are artefacts of the DFT of a finite length signal.}
\label{Fig:3}
\end{figure}

Off-resonance there is incomplete population transfer between $\ket{b}$ and $\ket{a}$. The resulting spectrum is given in Appendix A and  illustrated by Fig.\ 2(d). From this we observe:

\begin{enumerate}
\item{The center of the peaks is shifted by $|\Delta\omega/2|$ in each direction, depending on the peak group.}
\item{The splitting of peaks within a group increases from $2\varpi$ to $\sqrt{(2\varpi)^2 + (\Delta\omega)^2}$.}
\item{The intensity of the coherence peaks (i.e., AB and BA) relative to the dominant population peak (i.e., BB) is proportional to $2\varpi/\Omega$, and is a  measure of the strength of the coherences. Similarly,  the intensity of the population peak AA  relative to  BB is proportional to $4(\varpi/\Omega)^2$, and is a measure of the maximum population transfer between $\ket{b}$ and $\ket{a}$.}
\item{All but one of the peaks in group BB vanish for off-resonance if $2\varpi \ll\Omega$, as in this case no coherences between $\ket{b}$ and $\ket{a}$ are established.}
\end{enumerate}

Fig.\ \ref{Fig:3} shows a  calculated spectrum of the two-level system. The wavefunction overlap, $S(t_1,t_2,t_3)$, was performed numerically exactly via a Trotter decomposition of the evolution operator using the TNT Library\cite{Al-assam17}, while its Fourier transform, $\tilde{S}(\omega_1, t_2, \omega_3)$, was performed via a discrete Fourier transform. Clearly visible are the four groups of four peaks in accordance with the schematic picture shown in Fig.\ 2(d).
Fig.\ 2(d) and Fig.\ 3 illustrate the central result of our work, namely that our proposed protocol gives a very different 2D spectrum from the standard spectrum of coherence illustrated in Fig.\ 2(b).

\subsection{Noisy Environment}

\begin{figure}[h]
\centering
  \includegraphics[width=0.65\linewidth]{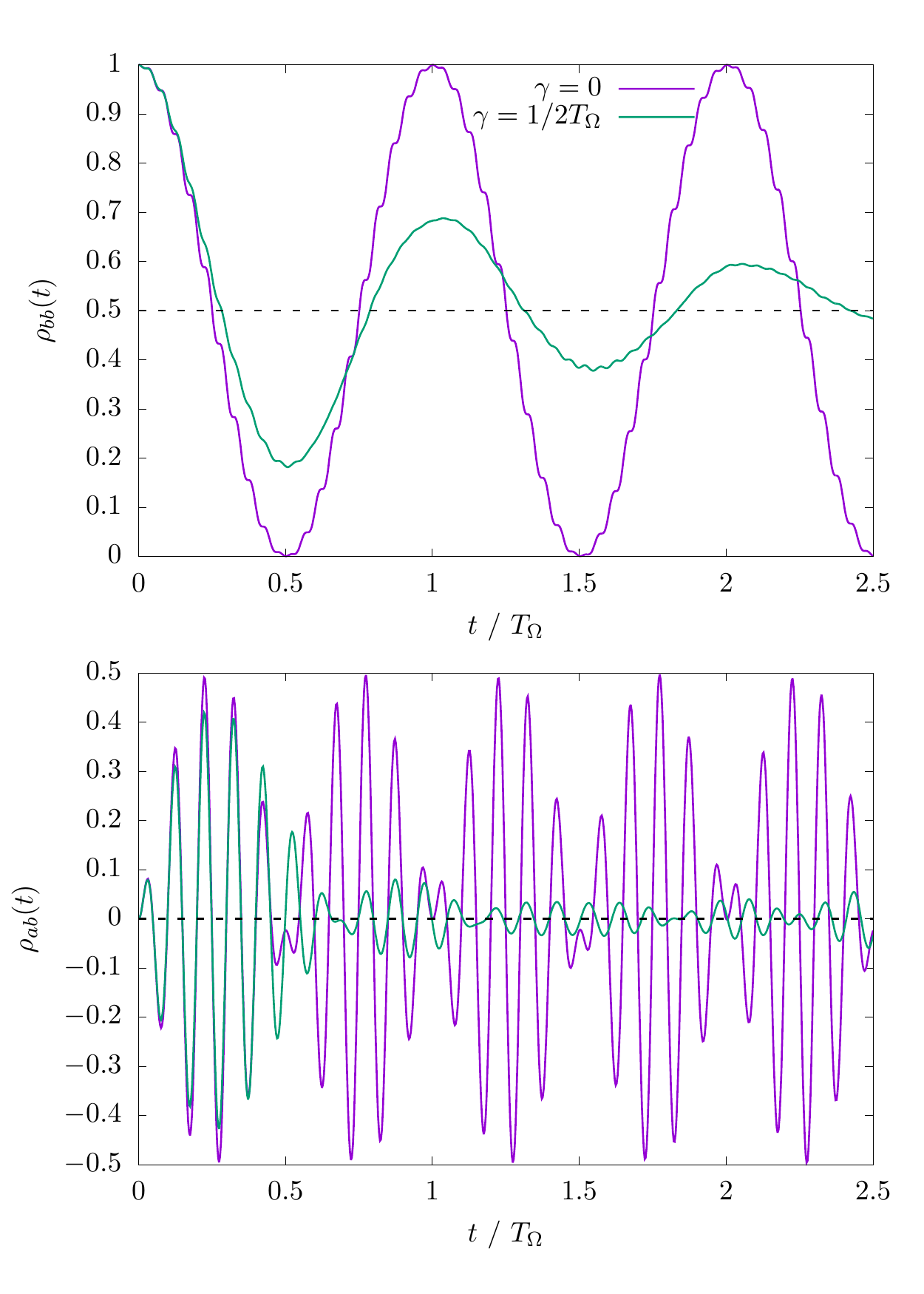}
  \caption{(a) The population of $\ket{b}$, $\rho_{bb}$,  as a function of time at resonance, $\omega = \omega_0$ (where $\rho_{aa} = 1-\rho_{bb}$).
  (b) The coherences between  $\ket{a}$ and $\ket{b}$, $\rho_{ab}$, as a function of time.
  The autocorrelation function of the system-bath interaction satisfies Eq.\ (\ref{Eq:4}). Purple (no dephasing), green (dephasing time $\gamma^{-1} = 2T_{\Omega}$, where $T_{\Omega}$ is the Rabi period). We note that in the case of no-dephasing $\rho_{bb}(t)$ deviates from $|\psi_b(t)|^2$ with $\psi_b(t)$ given by Eq.\ (\ref{Eq:9}) because this is a numerically exact solution of the two-level system and not the RWA.}\label{Fig:4}
\end{figure}

\begin{figure}[h]\label{Fig:5}
\includegraphics[width=0.7\linewidth]{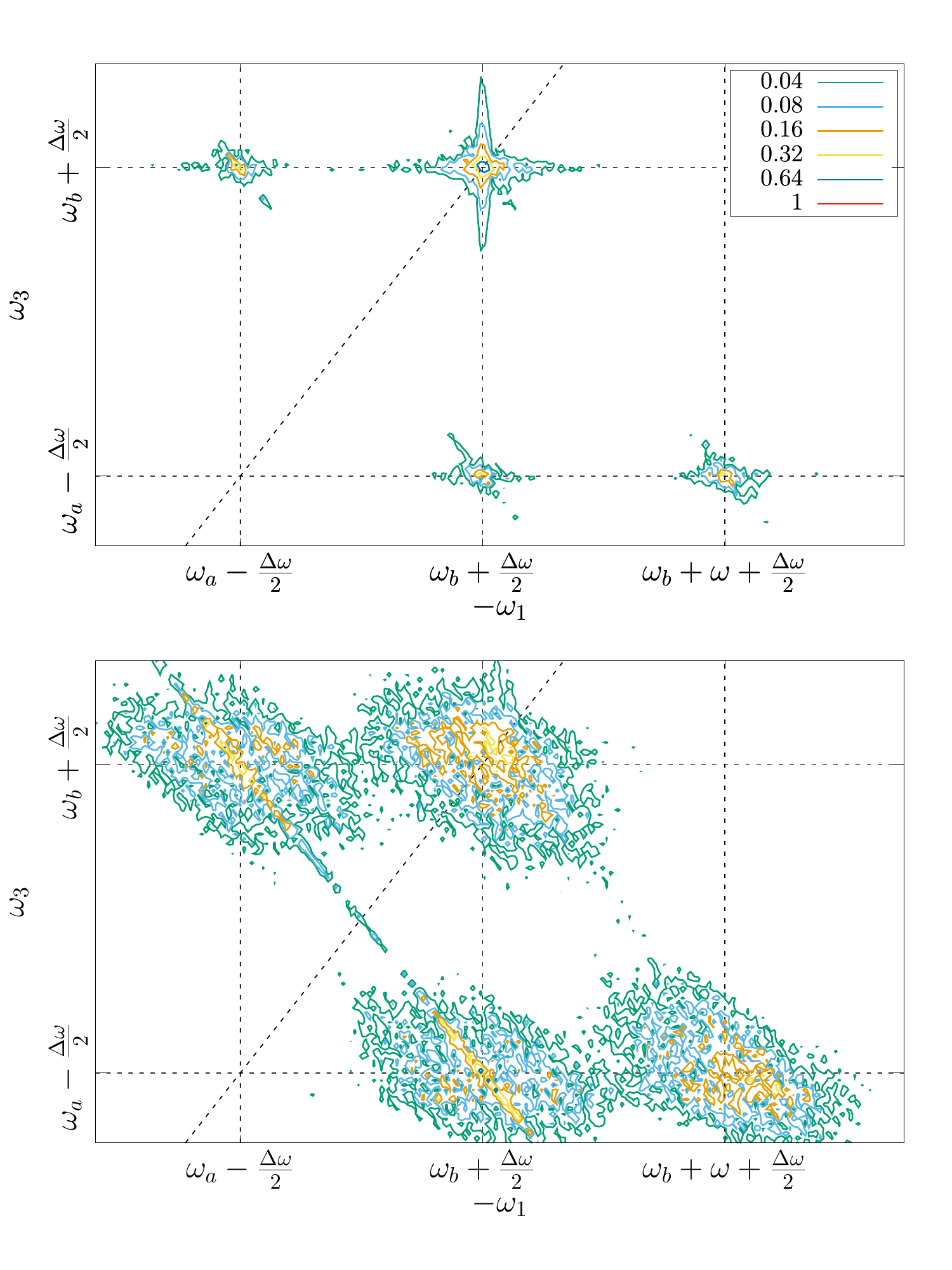}
\caption{Dephasing acting on the system and the effect on the spectra. The upper panel shows fast dephasing with $\gamma^{-1} = T_{\Omega}/4$, while the lower panel shows slower dephasing with $\gamma^{-1} = T_{\Omega}$, where $T_{\Omega}$ is the Rabi period. In both cases the fine structure of the peak groups is destroyed while the boost in $-\omega_1$ at $\omega_3=\omega_a-\Delta \omega/2$ is maintained.  $\Delta \omega/\omega = 0.9$, $\varpi/\omega_0 = 0.25$ and $t_2=0$.}
\end{figure}

Noise destroys coherences, so we now consider its role in changing our predicted spectra. We can incorporate noise into the interaction by randomising the phase. In particular, we choose $\phi(t)$ so that the autocorrelation function of the system-bath interaction satisfies
\begin{equation}\label{Eq:4}
  \langle V(t) V(0) \rangle  = |V(0)|^2\cos (\omega t) \exp(- \gamma t),
\end{equation}
where $\gamma$ is the dephasing rate. The form of Eq.\ (\ref{Eq:4}) is chosen so as to represent damped, harmonic system-bath interactions at ambient temperatures, e.g., torsional oscillations of  monomers subject to Brownian fluctuations.
Eq.\ (\ref{Eq:4}) is achieved if $\phi(t)$ in Eq.\ (\ref{Eq:3}) satisfies
\begin{equation}
\phi(t) = \int_0^t \delta\omega(t')\text{d}t',
\label{Eq:stoch1}
\end{equation}
where $\delta\omega(t)$ is a temporally-correlated noise function satisfying
\begin{equation}
\left\langle \delta\omega(t)\delta\omega(0)\right\rangle = (\delta\omega)^2\ex^{-t/\tau_c}
\label{Eq:stoch2}
\end{equation}
and $\delta\omega\tau_c \ll 1$.\cite{Kubo69, Hamm11}

The numerical simulations of the model with the stochastic interaction given by Eqs. (\ref{Eq:stoch1}) and (\ref{Eq:stoch2}) were performed via a Trotter decomposition of the evolution operator using the TNT Library\cite{Al-assam17}. By taking the ensemble average over 100 trajectories with different temporal disorder we construct the system's reduced density matrix.

The effect of a system-bath dephasing interaction is illustrated in  Fig.\ \ref{Fig:4}.  Fig.\ \ref{Fig:4}(a) shows that the populations of $\ket{a}$ and  $\ket{b}$
equilibrate to equal values, while  Fig.\ \ref{Fig:4}(b) shows  that the coherences between  $\ket{a}$ and $\ket{b}$
decay.

The effect of dephasing on the 2D-spectrum is shown in Fig.\ 5. For fast dephasing ($\gamma^{-1} = T_{\Omega}/4$, where $T_{\Omega}$ is the Rabi period), coherences  between $\ket{a}$ and $\ket{b}$ do not have time to become fully established, so a single population peak BB (at $\sim(-\omega_b, \omega_b)$) dominates the spectrum.
For intermediate dephasing ($\gamma^{-1} = T_{\Omega}$), however, coherences are established, but noise destroys the resolution of each of the four sub-peaks within a group of peaks.
Nevertheless, one of the characteristic fingerprints of induced dynamical coherences, namely the boost along $\omega_1$ at $\omega_3 = \omega_a-\Delta \omega/2$, is still evident.

\section{Realistic Model}\label{Se:4}

So far we have considered a theoretical two-level system and have shown how the dynamics of that system can be observed via 2D-spectrum. In this section we describe practical realizations of that system. In particular, we consider two examples of exciton dynamics in conjugated polymer systems, namely  (A), energy transfer between two chromophores and (B), energy relaxation on the same chromophore. In both cases we assume that the `external' interaction driving the dynamics is the damped torsional motion of the monomers. We also describe the electric field polarizations required to selectively excite and de-excite the eigenstates.

Here we are concerned with the lowest energy excited states of a chromophore. A convenient theory to describe these states is the Frenkel exciton model. We show in Appendix B how this model maps onto our two-level system with a time-dependent coupling of the eigenstates determined by the monomer rotations.


\subsection{Energy Transfer}\label{Se:4.2}

\begin{figure}[h]
\centering
\includegraphics[width=0.65\linewidth]{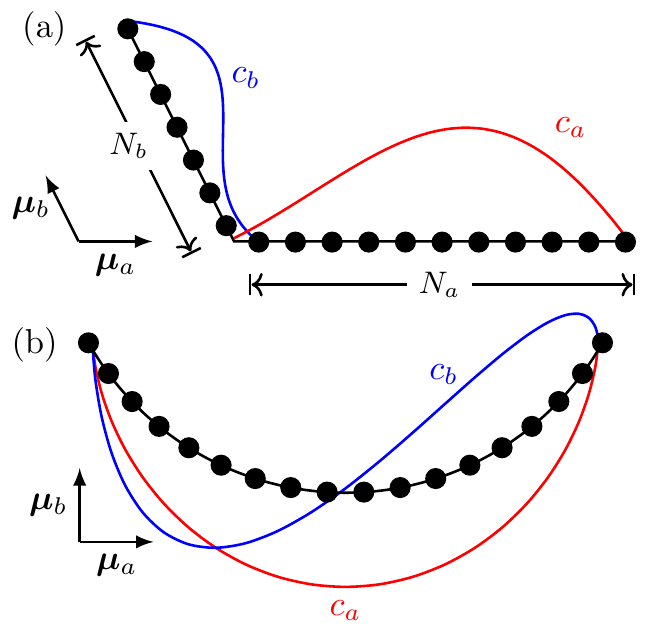}
\caption{A schematic of the  exciton center-of-mass wavefunctions $c_a$ (in red) and  $c_b$ (in blue) corresponding to the stationary states $\ket{a}$ and $\ket{b}$. (a)   neighboring chromophores of a polymer segment (represented by the dashed black curve), where $N_b < N_a$;
(b)   the same curved chromophore (represented by the dashed black curve).
$\bs{\mu}_a$ and $\bs{\mu}_b$ represent the transition dipole moments of these states. $N_j$ is the number of monomers in chromophore $j$.}\label{Fig:6}
\end{figure}

We first consider energy transfer between two almost orthogonal chromophores on a bent portion of a conjugated polymer, as shown in Fig.\ 6(a). In practice, chromophores are defined by the spatial extent of local exciton ground states (LEGSs), whose spatial extent is determined by Anderson localization of the exciton center-of-mass wavefunction\cite{Book}. The boundary between chromophores is determined by the spatial distribution of disorder and possible `conjugation breaks'. In all cases, the LEGSs overlap at the boundaries.

To achieve the protocol described in Section \ref{Sec:2.4}, exciton $\ket{b}$ is first excited, which means that the electric field is initially polarized along the transition dipole moment of $\ket{b}$. Thus, $ {\bf e} \propto \bm{\mu}_b$, and $\hat{O}_{\alpha} \propto \hat{\bf{\mu}}_b$ and $\hat{O}_{\beta} \propto \hat{\bf{\mu}}_b$.
The transfer of amplitude to exciton $\ket{a}$ is then determined by setting the electric field polarization along components of the transition dipole moments of both $\ket{a}$ and $\ket{b}$. Thus, $ {\bf e} \propto (\bm{\mu}_a + \bm{\mu}_b )$, and
$\hat{O}_{\gamma} \propto (\hat{\bf{\mu}}_a+\hat{\bf{\mu}}_b)/\sqrt{2}$ and $\hat{O}_{\delta} \propto (\hat{\bf{\mu}}_a+\hat{\bf{\mu}}_b)/\sqrt{2}$.

For a typical conjugated polymer at room temperature, e.g., poly(para-phenylene), $\bar{\delta \theta} \approx 0.1$ rads, $J \approx 2$ eV  and $ J_{SE}   \sin 2\theta \approx 1$ eV (see Appendix B). The wavefunction overlap is determined by disorder, but is typically $\sim 0.1$. Thus, $\hbar\varpi \lesssim 0.005 $ eV, implying at resonance a population transfer period   $T_{\Omega} \sim 1$ ps.
The torsion oscillation energy is $\hbar \omega = 0.02$ eV and thus  resonance is achieved if the difference in chromophore sizes satisfies\footnote{Because $E(N) = a -b/N^2$ for $N \gg 1$, where $a$ and $b$ are constants.} $\Delta N/N^3 \approx 5\times 10^{-4}$, e.g., $N_b = 18$ and $N_a = 22$. In this case $\varpi/\omega_0 = 0.25$, which corresponds to the parameters used for Fig.\ \ref{Fig:3}. In practice, noise will smear the spectrum, so a result more like Fig.\ 5 is expected.

\subsection{Energy Relaxation}\label{Se:4.3}

As a second example we consider energy relaxation from a higher to a lower energy exciton, as shown in Fig.\ 6(b). Here, $c_a$ is the LEGS while $c_b$ is a locally excited exciton state (LEES)\cite{Book}. Importantly, $c_b$ has one node so that on  a curved chromophore its  transition dipole moment is almost orthogonal to that of $c_a$\cite{Barford16a}. To achieve the protocol described in Section \ref{Sec:2.4},  $\hat{O}_{\alpha} \propto \hat{\bf{\mu}}_b$, $\hat{O}_{\delta} \propto \hat{\bf{\mu}}_b$, $\hat{O}_{\beta} \propto (\hat{\bf{\mu}}_a+\hat{\bf{\mu}}_b)/\sqrt{2}$ and $\hat{O}_{\gamma} \propto (\hat{\bf{\mu}}_a+\hat{\bf{\mu}}_b)/\sqrt{2}$.
In this case, resonance is achieved if the  chromophore size  $N \approx 50$, when again $\varpi/\omega_0 = 0.25$.

Before concluding this section, we note that the monomer rotations that cause interstate coupling also introduces a time-dependent diagonal term into the two-level model. However, because the angular frequency of this motion, i.e.,\ $\omega$, is so much smaller than $\omega_a$ and $\omega_b$, this term has a negligible effect on the 2D-spectra.


\section{Concluding Remarks}

We have proposed a  protocol for performing 2D-spectroscopy experiments that measure induced time-dependent coherences between the stationary states of a two-level system.
This protocol gives a rich 2D-spectrum characteristic of  dynamical coherences that differs from the observed coherences of the non-stationary state $\ket{\Psi(t=0)} = (\ket{a} + \ket{b})/\sqrt{2}$. The 2D-spectrum for  the non-stationary state  consists of populations at $(-\omega_a,\omega_a)$ and $(-\omega_b,\omega_b)$, and coherences  at $(-\omega_a,\omega_b)$ and $(-\omega_b,\omega_a)$. In contrast, the 2D-spectrum for measuring induced dynamical coherences proposed here differs in two distinct ways. First,  each peak splits into a  group of four peaks,  with the splitting of the peaks determined by the population transfer (Rabi) period, $T_{\Omega}$.
Second, at $\omega_3  \sim \omega_a$ there is a boost along the $\omega_1$ axis by $\omega$, the angular frequency of the system-bath interaction that induces these dynamics. These two features imply that this proposed protocol provides a unique fingerprint for induced dynamical coherences and the system-bath interactions.
However, for weak system-bath interactions  coherences are only established at or close to resonance, i.e., $\omega \simeq \omega_0$. Furthermore, coherences are destroyed by strong dephasing, meaning that only the population  at $(-\omega_b,\omega_b)$ would be observed.

We illustrated this approach using one of the `rephasing' (or photon-echo) diagrams in 2D-spectroscopy (i.e., the stimulated emission diagram, where $t_{\alpha} \leq t_{\beta} \leq t_{\gamma}$). However, the ground state bleach diagram (where  $t_{\gamma} \leq t_{\beta} \leq t_{\delta}$)\footnote{This diagram is labeled `C' in Ref.\cite{AHMarcus07} and $R_2$ in Ref.\cite{Hamm11}.} is also equally appropriate, as in both of these diagrams the first and third order trajectories evolve in the excited state manifold during either $t_1$ or $t_3$.

We next discussed  exemplars of this model, namely a conjugated polymer whose torsional fluctuations act as the time-dependent interaction. The two stationary states are either excitons localized on neighboring chromophores, or the two lowest excited states on the same chromophore. Since both the angular frequency of the torsional oscillations and the strength of the interaction that drives the dynamics is small (i.e., $\sim 0.01 - 0.02$ eV) it is important that the system is on or close to resonance for coherences to be induced and observed via 2D-spectroscopy. In addition, since these energy scales are much smaller than other possible electronic sources of homogeneous and inhomogenous broadening, e.g., by fast density fluctuations, we might expect that our example is only relevant for a polymer in an inert environment not subject to  sources of large dynamical and static disorder.

We hope that this proposed technique will have applications in other fields of molecular and condensed matter physics in the study of quantum coherences.

\begin{acknowledgments}
This work was performed using the Tensor Network Theory Library, Beta Version 1.2.1 (2016), S. Al-Assam, S. R. Clark, D. Jaksch, and the TNT Development team, www.tensornetworktheory.org.
\end{acknowledgments}


\vfill\pagebreak

\appendix
\section{Analytical Results}\label{Se:3.1}

Since the wavefunction overlap, $\langle\Psi_{\alpha\beta\gamma} |\Psi_{\delta}\rangle $, obtained via the RWA (i.e., Eq.\ (\ref{Eq:8}) and (\ref{Eq:9})) is a sum of a product of complex exponentials, its Fourier transform with respect to $t_1$ and $t_3$ results in a sum of a product of delta-functions. In particular, $\tilde{S}(\omega_1, t_2, \omega_3)$ corresponds to four groups of four-peaks. Referring to Fig.\ (2), we identify the four groups as follows.
\begin{itemize}
\item{
The group AA are associated with the population of $\ket{a}$:
\begin{widetext}
\begin{eqnarray}
\nonumber
\tilde{S}_{AA}    && =   \left(\frac{\varpi}{\Omega} \right)^2
  [ \delta(\omega_{1}+(\omega_a+\omega - (\Delta \omega \pm \Omega)/2))\delta(\omega_{3}-(\omega_a-(\Delta \omega \pm \Omega)/2))   \\
  \nonumber
  && - \delta(\omega_{1}+(\omega_a+\omega-(\Delta \omega \pm \Omega)/2))\delta(\omega_{3}-(\omega_a-(\Delta \omega \mp \Omega)/2)) \exp(\pm\textrm{i}\Omega t_2) ].\\
\end{eqnarray}
\end{widetext}
}
\item{
The group BB are associated with the population of $\ket{b}$:
\begin{widetext}
\begin{eqnarray}\label{Eq:16}
\nonumber
 \tilde{S}_{BB} &&  =   \frac{1}{4} \left( 1 \mp  \frac{\Delta \omega}{\Omega} \right)^2
  \delta(\omega_{1}+(\omega_b+(\Delta \omega \pm \Omega)/2))\delta(\omega_{3}-(\omega_b+(\Delta \omega \pm \Omega)/2))   \\
  \nonumber
  && +
  \frac{1}{4} \left( 1 -  \left(\frac{\Delta \omega}{\Omega}\right)^2\right) \delta(\omega_{1}+(\omega_b+(\Delta \omega \pm \Omega)/2))\delta(\omega_{3}-(\omega_b+(\Delta \omega \mp \Omega)/2)) \exp(\pm\textrm{i}\Omega t_2),\\
\end{eqnarray}
\end{widetext}
where $\Omega$ is given by Eq.\ (\ref{Eq:11}).
}
\item{
The group AB are associated with coherences between $\ket{a}$ and $\ket{b}$:
\begin{widetext}
\begin{eqnarray}
\nonumber
\tilde{S}_{AB}  && =     \mp \frac{\varpi}{2 \Omega} \left(1 \mp \frac{\Delta \omega}{\Omega}\right)
  \delta(\omega_{1}+(\omega_a-(\Delta \omega \mp \Omega)/2))\delta(\omega_{3}-(\omega_b+(\Delta \omega \pm \Omega)/2)) \exp(-\textrm{i}\omega t_2)
   \\
  \nonumber
  && \mp
  \frac{\varpi}{2 \Omega} \left(1 + \frac{\Delta \omega}{\Omega}\right)
 \delta(\omega_{1}+(\omega_a-(\Delta \omega \pm \Omega)/2))\delta(\omega_{3}-(\omega_b+(\Delta \omega \mp \Omega)/2))
 \exp(-\textrm{i}(\omega \mp \Omega)t_2). \\
\end{eqnarray}
\end{widetext}
}
\item{
Similarly, the group BA are associated with coherences between $\ket{a}$ and $\ket{b}$:
\begin{widetext}
\begin{eqnarray}\label{Eq:19}
\nonumber
 \tilde{S}_{BA}  && =  \mp \frac{\varpi}{2\Omega} \left(1 \mp \frac{\Delta \omega}{\Omega}\right)
  \delta(\omega_{1}+(\omega_b+\omega+(\Delta \omega \pm \Omega)/2))\delta(\omega_{3}-(\omega_a-(\Delta \omega \pm \Omega)/2)) \exp(\textrm{i}\omega t_2)
   \\
  \nonumber
  && \mp
  \frac{\varpi}{2 \Omega} \left(1 + \frac{\Delta \omega}{\Omega}\right)
 \delta(\omega_{1}+(\omega_b+\omega+(\Delta \omega \pm \Omega)/2))\delta(\omega_{3}-(\omega_a-(\Delta \omega \mp \Omega)/2))
 \exp(\textrm{i}(\omega \mp \Omega)t_2). \\
\end{eqnarray}
\end{widetext}
}
\end{itemize}

\section{Derivation of the Two-State Model for Conjugated Polymers}\label{Se:4.1}

The dynamics of Frenkel excitons in conjugated polymers is described by the Frenkel exciton model\cite{Book}, defined by
\begin{equation}\label{}
  \hat{H} = \sum_{n=1}^N \epsilon_ n \ket{n}\bra{n} +  \sum_{n=1}^{N-1} J_n(t) \left(\ket{n}\bra{n+1} + \ket{n+1}\bra{n}\right),
\end{equation}
where $n$ labels a monomer and $\epsilon_ n$ is the excitation energy of a monomer. $J_n$ is the exciton transfer integral,
\begin{equation}\label{Eq:23}
 J_ n(t) =   J_{DD} + J_{SE} \cos^2 \theta_n(t),
\end{equation}
where $J_{DD}$ is the through-space, dipole-dipole contribution, $J_{SE}\cos^2 \theta_n(t)$ is the through-bond, superexchange contribution, and $\theta_n$ is the dihedral angle between monomers. In general, $\theta_n(t) = \theta + \delta \theta_n(t)$, where the dynamical component  $\delta \theta_n(t)$ arises from thermal fluctuations.

Assuming that $\delta \theta_n(t) \ll \theta $ we can linearize Eq.\ (\ref{Eq:23}) to obtain
\begin{equation}\label{}
 J_ n(t) = J - J_{SE} \delta \theta_n(t) \sin 2\theta,
\end{equation}
where
\begin{equation}\label{}
 J =   J_{DD} + J_{SE} \cos^2 \theta.
\end{equation}
This linearization allows us to partition the Hamiltonian into a time-independent (system) part,
\begin{equation}\label{}
  \hat{H}_S = \sum_{n=1}^N \epsilon_ n \ket{n}\bra{n} +  \sum_{n=1}^{N-1} J \left(\ket{n}\bra{n+1} + \ket{n+1}\bra{n}\right),
\end{equation}
and a time-dependent (system-bath) part,

\begin{equation}\label{}
  \hat{H}_{SB} =  -  J_{SE} \sin 2\theta  \sum_{n=1}^{N-1} \delta \theta_n(t)\left(\ket{n}\bra{n+1} + \ket{n+1}\bra{n}\right).
\end{equation}

Transforming into the eigenkets of $\hat{H}_S$, i.e.,
\begin{equation}\label{}
\ket{j} = \sum_{n=1}^N c_{nj} \ket{n}
\end{equation}
(where $c_{nj}$ is the exciton wavefunction, i.e., $j=a,b$, shown in Fig.\ 6)
we have the system Hamiltonian
\begin{equation}\label{Eq:29}
  \hat{H}_S = \sum_{j=1}^N \hbar \omega_j \ket{j}\bra{j},
\end{equation}
\begin{equation}\label{}
  \hbar \omega_j =   2J \sum_n c_{n,j}c_{n+1,j}.
\end{equation}
Similarly, the system-bath Hamiltonian is
\begin{equation}\label{Eq:31}
  \hat{H}_{SB} = \sum_{ij} V_{ij}(t)\ket{i}\bra{j},
\end{equation}
where
\begin{equation}\label{}
 V_{ij}(t) = -  J_{SE} \sin 2\theta  \sum_{n=1}^{N-1} \delta \theta_n(t)  \left(c_{n,i}c_{n+1,j}+c_{n,j}c_{n+1,i}\right).
\end{equation}

For a polymer  subject to stochastic thermal fluctuations,  the dynamical component of the torsional angle satisfies
\begin{equation}\label{}
 \delta\theta_n(t) =  \bar{\delta\theta}\cos(\omega t + \phi_n(t)),
\end{equation}
where $\bar{\delta\theta} = \left(k_B T/K_{\textrm{rot}}\right)^{1/2}$, $\omega$ is the rotational angular frequency and $K_{\textrm{rot}}$ is the rotational force constant.

Finally, by only retaining two states, $\ket{a}$ and $\ket{b}$, this model maps onto the two-level model  introduced  Section \ref{Se:2.1} and we find that the interaction strength is,
\begin{equation}\label{Eq:34}
2\hbar\varpi  = - J_{SE}  \bar{\delta\theta} \sin 2\theta \sum_{n=1}^{N-1}   \left(c_{n,a}c_{n+1,b}+c_{n,b}c_{n+1,a}\right).
\end{equation}

\bibliography{references}

\end{document}